 \documentclass[preprint,aps,prd,showpacs,preprintnumbers,amsmath,amssymb,amsfonts, nofootinbib,a4paper,12pt]{revtex4-1}

\usepackage{hyperref}

 \usepackage{enumerate}
 \usepackage{graphicx} 
 \numberwithin{equation}{section}

\def\stamp{--- {\bf \today} --- {\bf \jobname.tex}}

\def\cA{\mathcal{A}}

\def\stamp{--- {\bf \today} --- {\bf \jobname.tex}}

\def\sign(#1){\textrm{sign}(#1)}

\def\cA{\mathcal{A}}

\def\cnull{} 
\def\BE{\begin{equation}} 
\def\EE{\end{equation}}

 \def\<#1|#2){\left\langle#1|#2\right\rangle} 
 \def\<#1|#2|#3]{\left\langle#1|#2|#3\right ]}

\def\an[#1,#2]{\left\langle#1\,#2\right\rangle} 
\def\aq[#1,#2,#3]{\left\langle#1|#2|#3\right]} 
\def\qa[#1,#2,#3]{\left[#1|#2|#3\right\rangle} 
\def\sq[#1,#2]{\left[#1\,#2\right]} 
\def\spa#1.#2{\left\langle#1\,#2\right\rangle} 
\def\spab[#1,#2,#3]{\left\langle#1|#2|#3\right]} 
\def\spba[#1,#2,#3]{\left[#1|#2|#3\right\rangle} 
\def\spb#1.#2{\left[#1\,#2\right]} 
\def\lor#1.#2{\left(#1\,#2\right)}

\begin{document}
\preprint{IHES/P/14/11, IPTH-t14/030} 
\title{Scattering Equations and String Theory Amplitudes} 
 
\author{\vspace{.5cm}{\bf
N.~E.~J~Bjerrum-Bohr${}^a$, 
P.~H.~Damgaard${}^{a}$, 
P.~Tourkine$^{b}$  and  
P.~Vanhove$^{b,c}$\\ \vskip0.3cm ${}^a$  }
Niels Bohr International Academy and Discovery Center,\\ 
The Niels Bohr Institute, Blegdamsvej 17, 
DK-2100 Copenhagen \O, Denmark,\\ \vskip0.2cm 
${}^b$ CEA, DSM, Institut de Physique Th{\'e}orique, IPhT, CNRS, MPPU,\\ 
URA2306, Saclay, F-91191 Gif-sur-Yvette, France\\ \vskip0.2cm 
${}^c$ Institut des Hautes Études Scientifiques, F-91440,  Bures sur Yvette, France
\\ \vskip0.4cm 
{\bf Email:} bjbohr@nbi.dk, phdamg@nbi.dk, piotr.tourkine@cea.fr, pierre.vanhove@cea.fr}

\begin{abstract}Scattering equations for tree-level amplitudes are
  viewed in the context of string theory.  To this end we are led to define a new dual model 
whose amplitudes coincide with string theory in both the small and large $\alpha'$ limit, computed 
algebraically on the surface of solutions to the scattering equations. 
Because it has support only on the scattering equations, it can be solved exactly, 
yielding a simple resummed model for $\alpha'$-corrections to all orders. 
We use the same idea to generalize scattering equations to amplitudes with fermions
and any mixture of scalars, gluons and fermions. In all cases checked we find
exact agreement with known results.
\end{abstract}
 

\maketitle 
 

\section{Introduction}\label{sec:introduction} 
In a series of remarkable papers, Cachazo, He and Yuan (CHY) have proposed that  
tree level scattering of massless particles in any dimension can be constructed 
from algebraic solutions of a set of kinematic scattering equations~\cite{Cachazo:2013gna, 
Cachazo:2013hca,Cachazo:2013iea}. This idea had originally been supported by a number of  
highly non-trivial observations and checks, 
and also by explicit amplitude computations for a large number of 
external legs. A proof of this surprising construction has recently been provided 
for scalar amplitudes and gluon amplitudes by Dolan and Goddard in ref.~\cite{Dolan:2013isa}  
based on Britto-Cachazo-Feng-Witten (BCFW)~\cite{Britto:2005fq} recursion. These authors have also shown how to generalize 
the construction to massive scalars, extending the specific construction for scalars 
of ref.~\cite{Cachazo:2013iea} to any theory of scalars with only 3-point vertices, again 
in any dimension.  
 
The whole setup of the scattering equation approach is eerily
reminiscent of string theory, and indeed it was recognized early
on~\cite{Cachazo:2013hca} that these scattering equations coincide
with the saddle point equations of the Gross-Mende
limit~\cite{Gross:1987ar}.  But this also represents a conundrum: The
Gross-Mende limit is that of high-energy scattering of strings
corresponding to $\alpha' \to \infty$, not the opposite limit of
$\alpha' \to 0$ where the field theory of pointlike particles
emerges. Indeed, in the $\alpha' \to 0$ limit of string theory an
entirely different formalism arises, even though, eventually, the same
tree-level amplitudes come out. It is as if the scattering equation
approach has managed to obtain a different limit of $\alpha' \to 0$,
while retaining aspects of high-energy scattering of strings.  In the
twistor string frameworks~\cite{Mason:2013sva,Berkovits:2013xba,
  Adamo:2013tsa,Gomez:2013wza}, it has been demonstrated that one can
naturally impose the scattering equations in an alternative path
integral formulation. Many other indications of a close connection to
string theory can be found. In~\cite{Cachazo:2013gna} it was thus
shown that the scattering equations are intimately related to the
momentum kernel $S$~\cite{BjerrumBohr:2010ta,BjerrumBohr:2010hn}
between gauge and gravity theories (and hence between open and closed
strings). Similarly, scattering equations manifestly operate with a
basis of $(N-3)!$ amplitudes, in agreement with what is inferred from
Bern-Carrasco-Johansson (BCJ) relations~\cite{Bern:2008qj} and that
follows directly from string
theory~\cite{BjerrumBohr:2009rd,Stieberger:2009hq}
(see~\cite{Naculich:2014rta,Naculich:2014naa,Schwab:2014fia} for
applications to massless and massive amplitude). A more direct link
between BCJ relations and scattering equations has also been
proposed~\cite{Cachazo:2013iea,Litsey:2013jfa,Monteiro:2013rya}.
Finally, some algebraic relations arising in the string theory
computation of disk
amplitudes~\cite{Mafra:2011nv,Mafra:2011nw,Broedel:2013tta,Barreiro:2013dpa}
have also found use in the scattering equation formalism. All of these
examples indicate a close connection to string theory.

In this paper we suggest a new dual model that gives field theory amplitudes back in the 
$\alpha' \to 0$ limit and that, when $\alpha' \to \infty$, is governed by 
the Gross-Mende saddle point of high-energy string scattering:
\begin{multline} 
{\frak A}_N  
 = \int \, \prod_{i=2}^{N-2} dz_i 
\prod_{1\leq i< j\leq N} |z_i-z_j|^{2\alpha' k_i\cdot k_j}\times\cr
\times{(z_1-z_{N-1})^2(z_{N-1}-z_N)^2(z_N-z_1)^2\over \prod_{i=1}^{N} (z_i - z_{i+1})^{2}}
\prod_{i\neq 1,N-1,N}\delta(S_i)\,,  
\label{scalarnewdual1}
\end{multline}
where the integration is over the ordered set $z_1<z_2<\cdots<z_N$ and the
three points $z_1$, $z_{N-1}$ and $z_N$ have been fixed by
$SL(2,\mathbb C)$ invariance.
This expression differs from the CHY prescription by the Koba-Nielsen
factor $\prod_{1\leq i< j\leq N-2} |z_i-z_j|^{2\alpha' k_i\cdot k_j}$,
and differs from the usual string theory amplitude prescription by the
delta function constraints
$(z_1-z_{N-1})(z_{N-1}-z_N)(z_N-z_1)\,\prod_{i\neq 1,N-1,N}\delta(S_i)\prod_{i=2}^{N-2} (z_i - z_{i+1})^{-1}$.

At intermediate values of $\alpha'$, because of the
  delta function constraint, these tree-level amplitudes differ from the ones 
evaluated in the Ramond-Neveu-Schwarz (RNS) formalism or the pure
  spinor formulation~\cite{Berkovits:2001us}. The difference with the traditional string
  theory tree-level amplitude is discussed in section~\ref{sec:IBP},
  where we show that  the  above prescription has a soft high-energy
  behavior similar to the one of the conventional string theory. Therefore the
  prescription retains some fundamental properties of stringy amplitudes.
It would be interesting to relate the prescription given in this paper
to an $\alpha'$-extension  of Berkovits'
modified pure spinor prescription in the infinite tension limit~\cite{Berkovits:2013xba}. 
 We view it as a new dual model that could
have been introduced long ago. Indeed, the approach by Fairlie et
al.~\cite{Fairlie:1972zz} (reviewed in~\cite{Fairlie:2008dg}) by
imposing on a scalar dual model a minimal area constraint is closely
related to this, only missing the more general context and the new
connection to the field theory limit $\alpha' \to 0$ that we provide
here. 
 
The connection with the usual quantum field theory limit of string theory
and its high-energy limit is summarized in the following diagram
showing that new amplitude $\mathfrak A_N$ interpolates between the
CHY prescription and a high-energy limit with the Gross-Mende saddle point. 
$$
\includegraphics[height=5cm]{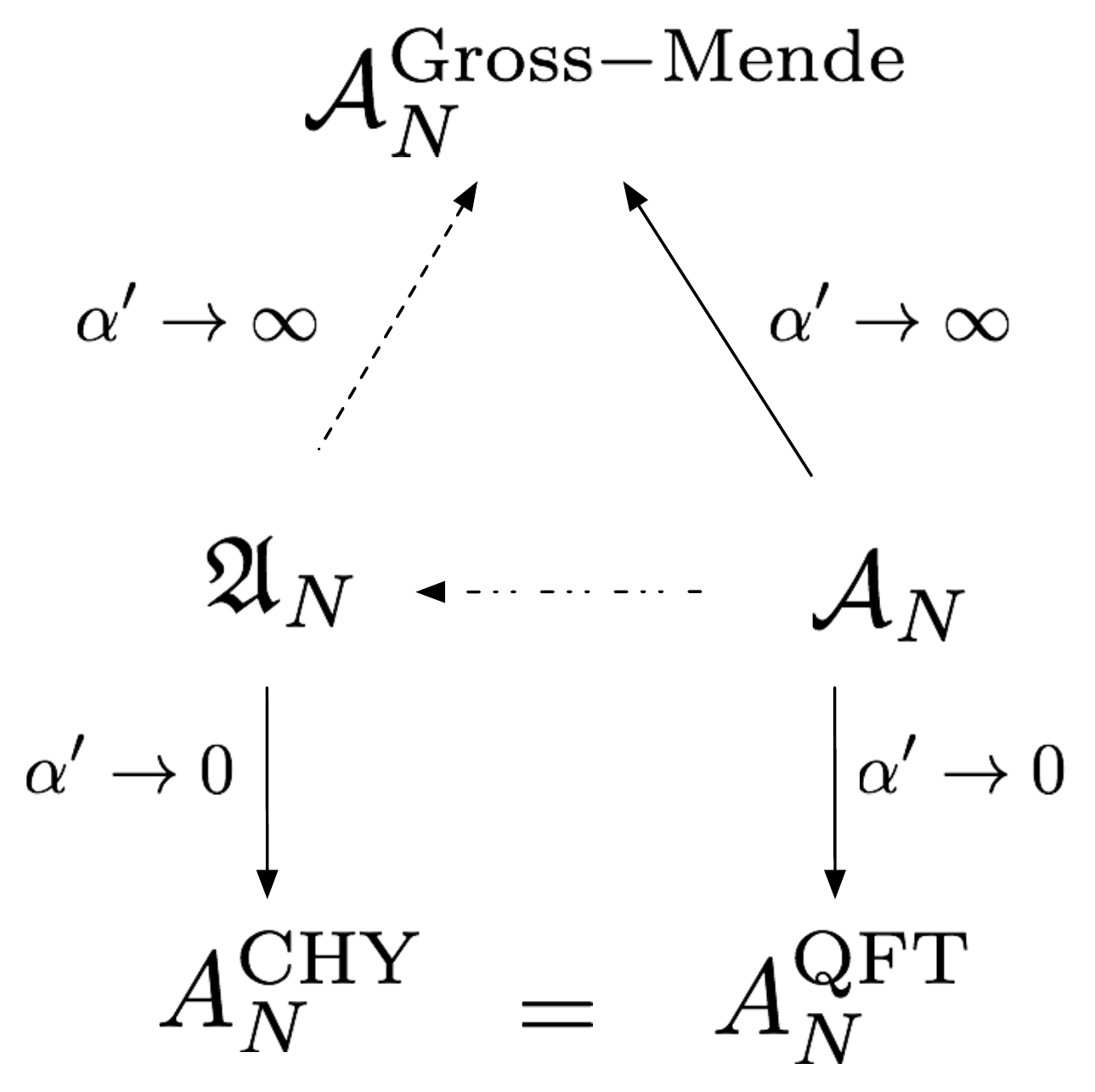}
$$
To show that the approach we suggest here also holds in a broader context
  than the original CHY prescription, we illustrate how the prescription
in~\eqref{scalarnewdual1} can be extended to include fermions as in
the superstring. We demonstrate explicitly that this produces correct
amplitudes with fermions in a few simple cases. Also examples of mixed
amplitudes with scalars, gluons and fermions will be considered and shown to agree
with known results. 

Our paper is organized in the following way. First, in section~\ref{sec:SQeq}, we briefly review the scattering equations 
and their solution in the field theory limit. Next, we motivate the 
simple new dual model of scalars in~\eqref{scalarnewdual1}. By imposing on the integrand the scattering equations, we obtain a simple 
scalar analog of the general framework of this paper: a model that reproduces the field 
theory limit on the surface of solutions to the scattering equations as $\alpha' \to 0$ and 
which reproduces the Gross-Mende solution in the limit of $\alpha' \to
\infty$. Then in section~\ref{sec:gauge}, we consider the case of  amplitudes involving gauge fields,
by  first briefly recalling how to compute the corresponding gluon amplitudes  
in string theory. The expression for the string integrand is rather cumbersome, but it 
can be rearranged into a form identical to the Pfaffian prescription of refs.~\cite{Cachazo:2013gna, 
Cachazo:2013hca,Cachazo:2013iea}, up to additional terms that formally are suppressed as 
$1/\alpha'$. 
We show  that all the additional pieces are proportional to the scattering 
equations after suitable integrations by parts manipulations familiar
in string theory (see~\cite[eqs.~(6.2.25)]{Polchinski:1998rq}
and~\cite{Stieberger:2007jv,Mafra:2011nv,Mafra:2011nw,Broedel:2013tta,Barreiro:2013dpa,Stieberger:2014hba}). 
Therefore, on the surface of solutions to these 
equations they do not contribute,  and the resulting modified integrand for our dual model in~\eqref{scalarnewdual1} yields the CHY amplitude prescription
in the field theory limit $\alpha' \to 0$.
In section~\ref{sec:IBP} we use this observation to show how to extend the
scalar dual model prescription to include gauge fields. 
This elementary construction is particularly easily understood in the case of the four-point gluon  
amplitude. We also show how such manipulations extend to higher point 
amplitudes. 
Finally, in section~\ref{sec:fermions} we discuss how to extend these considerations to compute amplitudes with 
external fermions on the basis of the scattering equations, and how mixed amplitudes with scalars,
fermions and vectors can be computed as well. We end with an outlook for future work.  

\section{Scattering Equations and a Dual Model Extension}\label{sec:SQeq} 
For scalar theories, the prescription given by the CHY prescription
for computing  $N$-point scalar 
amplitudes reads
\begin{equation} 
 A_{N\, \rm scalar}^{\rm CHY}= \int \, \prod_{i\neq 1,N-1,N}
\delta(S_i) \, {(z_1-z_{N-1})^2(z_{N-1}-z_N)^2(z_N-z_1)^2\over\prod_{i=1}^{N} (z_i-z_{i+1})^2}
 \, \prod_{i=2}^{N-2} dz_i\,,
\label{scalarCHY}
\end{equation} 
where the legs are ordered canonically from $1$ to $N$, and the
notation is such that $z_{N+1} \equiv z_1$.
Here $S_i$ denotes the $i$th scattering equation 
\begin{equation} 
S_i= \sum_{j\neq i} {k_i\cdot k_j\over z_i-z_j} \; =\;0\,. \label{sceq}
\end{equation} 
In the following we will fix the three points  $z_1=0$, $z_{N-1}=1$ and $z_N=\infty$. 
The CHY  prescription given  for computing $N$-point 
gauge theory amplitudes reads 
\begin{equation} 
 A_{N\,\rm gauge}^{\rm CHY}= \int \,{\rm Pf'}
 \Psi_N(z_i)\,\prod_{i\neq 1,N-1,N} 
\delta(S_i) \,{(z_1-z_{N-1})^2(z_{N-1}-z_N)^2(z_N-z_1)^2\over \prod_{i=1}^{N} (z_i-z_{i+1})}\,
\prod_{i=2}^{N-2} dz_i\,.
\end{equation} 
The function ${\rm Pf}'\Psi_N(z_i)$ is the {reduced} Pfaffian, ${\rm
  Pf}'(\Psi)_N(z_i)$, given by
\begin{equation} 
  {\rm Pf}'\Psi_N(z_i)=\frac{(-1)^{i+j}}{z_i-z_j}{\rm Pf}(
\Psi_{ij}^{ij})\,, 
  \label{e:pfp-def} 
\end{equation} 
where $\Psi_{ij}^{ij}$ is the matrix obtained from $\Psi$ by removing the rows 
and columns $i$ and $j$ (two rows and two columns removed). 
Gauge theory amplitudes  are obtained with
\begin{equation} 
 \Psi_N(z_i) =\begin{pmatrix} 
         A & -C^T \\ 
         C & B  
        \end{pmatrix}\,, 
\end{equation} 
where 
\begin{equation}  
A_{i,j}=\begin{cases} \displaystyle 
\frac{k_i\cdot k_j}{z_{i}-z_{j}} & i\neq j,\\ 
\displaystyle \quad ~~ 0 & i=j,\end{cases}  
~~  
B_{i,j} = 
\begin{cases} \displaystyle \frac{\epsilon_i\cdot \epsilon_j}{z_{i}-z_{j}} &
i\neq j,\\ 
\displaystyle \quad ~~ 0 & i=j,\end{cases}  
~~  
C_{i,j}=\begin{cases} 
\displaystyle \frac{\epsilon_i\cdot k_j}{z_i-z_j} & i\neq j\,,\\ 
\displaystyle -\sum_{l\neq i}\frac{\epsilon_i\cdot k_l}{z_i-z_{l}} &  
i=j\,.\end{cases} 
\label{e:pfaff-gauge}
\end{equation} 

Let us now try to see this construction in the light of old-fashioned dual models
with a dimensionful parameter $\alpha'$.
A simple dual model that yields the same massless scalar scattering amplitudes in the
limit $\alpha' \to 0$ is the following:
\begin{equation} 
{\cal A}_N \!=\left(g_o\over\sqrt{\alpha'}\right)^{N-2}\,{\alpha'}^{N-3}\,
\int\prod_{i=2}^{N-2}
dz_i\,{(z_1 - z_{N-1})(z_{N-1}-z_N)(z_N - z_1)\over \prod_{i=1}^N (z_i - z_{i+1})}
\prod_{1\leq i< j\leq N} |z_i-z_j|^{2\alpha' k_i\cdot k_j}\!
\,,  
\label{scalardual}
\end{equation}
where the integration is ordered along the real axis and $g_o$ is the
open string coupling constant.

Note how different the integration prescription is in the two cases. In the simple dual
model defined above, we integrate in an ordered manner along the real line after
having fixed again $z_1=0$, $z_{N-1}=1$ and $z_N=\infty$. In the integral defining
amplitudes based on scattering equations (\ref{scalarCHY}) 
the integral is saturated by the solutions to the delta function constraints.
This means that singularities that normally carry the whole amplitude in the $\alpha' \to 0$ limit
are harmless. 
Also the remaining part of the integrand is of course totally different, as
there is no trace of $\alpha'$ in (\ref{scalarCHY}). Yet, remarkably, for all $N$ the
$\alpha'\to 0$ limit of (\ref{scalardual}) yields exactly the same answer as  (\ref{scalarCHY}).
This suggests that it may be advantageous to view (\ref{scalarCHY}) as the leading term
of a more elaborate amplitude that depends on a parameter $\alpha'$.

Based on this perhaps rather na\"ive argument, let us introduce a very simple 
new dual model defined by amplitudes (using the relation $g_o=g_{\rm Yang\ Mills}\sqrt{2\alpha'}$ between the
open string coupling constant and the Yang-Mills (YM) coupling constant in ten dimensions)
\begin{equation} 
{\frak A}_N  
 =g_{\rm YM}^{N-2} \int  \prod_{i=2}^{N-2} dz_i 
\prod_{1\leq i< j\leq N} |z_i-z_j|^{2\alpha' k_i\cdot k_j}\!\!\!
\prod_{i\neq1,N-1,N}\!\!\!\delta(S_i) {(z_1-z_{N-1})^2(z_{N-1}-z_N)^2(z_N-z_1)^2\over
  \prod_{i=1}^{N} (z_i - z_{i+1})^2}\,. 
\label{scalarnewdual}
\end{equation}
Note that, effectively, this simply amounts to taking the dual model expression
and inserting the normalized delta function constraint\footnote{The
  delta function constraint has to be understood to include signs as in~\cite{Cachazo:2013hca}.
 In general, this can be given a 
precise interpretation in terms of contours in the complex plane via the global residue theorem \cite{Cachazo:2013hca,Dolan:2013isa}. However in all cases we have considered 
(even in the case of complex solutions to the scattering equations)
the na\"ive delta function constraint works as well, and of course the 
final result is real.}
\begin{equation}
{\alpha'}^{3-N} (z_1-z_{N-1})(z_{N-1}-z_N) (z_N-z_1)\prod_{i\neq1,N-1,N}\delta(S_i)\prod_{i=1}^{N} (z_i - z_{i+1})^{-1}\,,
\label{deltafunction}
\end{equation}
in the integrand. The overall powers  of $\alpha'$ can be understood
from the fact that it is natural from string theory to 
insert the delta function $\delta(\alpha'\,S_i)={\alpha'}^{-1}\, \delta(S_i)$.
Our claim is that this prescription, applied to 
open string theory amplitudes, provides a constructive way to reproduce
field theory amplitudes. In this expression one can set
$\alpha'$ to zero in the integrand to recover the CHY prescription.
 The justification of this point is the
subject of the next sections.

Massive scalar amplitudes can be dealt with easily, as they simply correspond
to replacing
\begin{equation}
\prod_{i=1}^N (z_i - z_{i+1})^{-1} ~\to \prod_{i=1}^N (z_i - z_{i+1})^{-1-\alpha' m^2}\,,
\end{equation}
in the integrand of (\ref{scalardual}). By differentiation of the integrand with
respect to $z_i$ we obtain the massive scattering equation proposed and proven 
to be correct in ref. \cite{Dolan:2013isa}. The fact that scattering equations arise
from differentiation with respect to the $z_i$ of external legs in the integrand
will play a crucial role in what follows.

In contrast to a more conventional dual model such as
(\ref{scalardual}), the new integral (\ref{scalarnewdual}) has a
totally smooth and finite limit $\alpha'\to 0$, where it of course
coincides with scalar field theory.  So has anything been achieved in
making such a trivial extension? A hint that this may be so is that in
the opposite limit $\alpha'\to \infty$, the amplitudes of
(\ref{scalarnewdual}) and (\ref{scalardual}), are both fixed by the
same Gross-Mende saddle point of high-energy string scattering.  So
this simple extension (\ref{scalarnewdual}) retains all the nice
properties of (\ref{scalarCHY}) when $\alpha' = 0$, and yields stringy
amplitudes in the opposite limit of $\alpha'\to \infty$. In between
these two limits we obviously have no immediate way to interpret the
amplitudes (\ref{scalarnewdual}), but these amplitudes are all
trivially computable due to the $\delta$-function constraint in the
measure.

What could be the meaning of the dimensional parameter $\alpha'$ here?
It would be tempting to view it as an inverse string tension. However,
such a point of view is not tenable. This becomes clear already in the
case of four-particle scattering, which has almost no resemblance at
finite $\alpha'$ to the corresponding Veneziano amplitude of
(\ref{scalardual}). There is not an infinite series of poles in the
amplitude that, rather, is more like that of ordinary field theory
with a trivial exponential damping factor. Indeed, because the limit
$\alpha' \to 0$ meets no singularity, amplitudes with either small or
large momenta can be found immediately at any value of $\alpha'$. At
$\alpha'=0$ the scattering amplitudes of (\ref{scalarnewdual}) are
just those of field theory, up to arbitrarily high energies. The
extension of (\ref{scalarCHY}) to the new dual model
(\ref{scalarnewdual}) looks much like dualized (color-ordered) scalar
field theory regularized with an ultraviolet cutoff
$1/\sqrt{\alpha'}$.

At this point, the dual model (\ref{scalarnewdual}) cannot be viewed
as anything else but a curiosity. If there is to be any substance in
it, and insight to be gained, we must see if a slightly more
sophisticated line of approach can yield new results. We therefore
turn to ordinary string theory, and explore the extent to which
similar considerations can be extended to massless gauge boson
scattering.

\section{Scattering Equations and Gauge Fields}\label{sec:gauge} 

 In this section we explore in some detail the properties of the
 prescription~\eqref{scalarnewdual}. It is well known that the requirement
of multilinearity in external polarization vectors conveniently can be implemented
in terms of auxiliary fermionic integrations in the string integrand. These real
Grassmann variables, when integrated out, produce a Pfaffian. This suggests that
the Pfaffian prescription of the previous section may be viewed as a remnant
of the string theory integrand, now only evaluated on the solutions to the scattering
equations. As we shall see, this is indeed the case. But instead of computing the
resulting Pfaffian directly, it is convenient to split it up into its separate
components, in this way illuminating which pieces give rise to the Pfaffian of
the previous section, and which do not.

%
\subsection{Multi-Pfaffian Structure of $N$-point Open-String Integrand.}

We first provide a new way to decompose the string theory integrand
for the scattering of $N$ gluons in the open superstring as a sum of
Pfaffians. This will include terms in the integrand of increasing powers of
$1/\alpha'$ as $N$ grows, but of course the full integral starts
with terms of order $1/\alpha'$ only. These terms of higher powers
of $1/\alpha'$ in the integrand can indeed be re-cast into terms 
that carry no explicit factor of $\alpha'$ by means of integrations
by parts. Such rewritings show that these terms do not contribute on
the surface of solutions to the scattering equations.  

In the RNS formalism, the vertex operators come in various
ghost pictures with respect to the superconformal ghost
$(\beta=\partial \xi \, e^{-\varphi}, \gamma=e^{\varphi})$~\cite{fms}. The
$-1$ ghost picture of the unintegrated vertex
operator for the emission of a gauge boson is then given by
\begin{equation}
  U^{(-1)}=g_o T^a: e^{-\varphi} \epsilon\cdot\psi  \, e^{ik\cdot X}:\,,
  \label{e:Uopen-1}
\end{equation}
while these in the $0$ ghost picture read
\begin{equation}
	U^{(0)}=g_o{\sqrt{2\over\alpha'}}T^a :(i\partial X^\mu + 2\alpha'
(k\cdot\psi) (\epsilon\cdot \psi))e^{ik\cdot X}:\,.
  \label{e:Uopen0}
\end{equation}
The corresponding integrated vertex operators are given by
\begin{equation}
\begin{split}
  &V^{(-1)}= \int dz : U^{(-1)}:  \,,\\
  &V^{(0)}= \int dz :U^{(0)}:\,.
  \end{split}
    \label{e:Vopen}
\end{equation}
The normalization of the operator-product expansion (OPE) on the boundary of the disk is such that 
\begin{eqnarray}
  X^\mu (z) X^\nu(0)&\simeq&-{\alpha'}\,\log|z|^2\,,\cr
  \psi^\mu(z)\psi^\nu(0)&\simeq&{\eta^{\mu\nu}\over z}\,,\\
\nonumber  e^{q_1\varphi(z)} e^{q_2\varphi(0)}&\simeq& {1\over z^{q_1q_2}}\,.  
\end{eqnarray}

At tree-level, to saturate the $+2$ background superghost charge, one
should set two vertex operators in the $-1$ ghost-picture, the rest can be
chosen in the $0$ ghost-picture. These two operators chosen in the $-1$ ghost picture,
for instance $V_1$ and $V_2$,
determine which lines and columns of the matrix one should remove to get the
correct reduced Pfaffian of equation~\eqref{e:pfp-def}. The $n$-gluon open-string amplitude $\cA_N^{}$ reads:
\begin{equation}
	{\cal A}^{}_N = {1\over \alpha'\,g_o^2}
		\langle cU^{(-1)}(z_1) cU^{(-1)}(z_{N-1}) cU^{(0)} (z_N)
		\int \prod_{i=2}^{N-2}d z_i 
		U^{(0)}(z_2)  \cdots U^{(0)}(z_{N-2}) \rangle \,.
\label{e:npt-open}
\end{equation}
where $g_o$ is the open-string coupling constant. A Pfaffian comes out of this integral simply because of the
Grassmann integral over a product of fermionic fields.

Focusing first on the purely fermionic part of
the correlator~\eqref{e:npt-open}, it involves a product of $2N-2$ fermionic
fields, among which $N-3$ are bilinears:
\begin{equation}
	\langle(\epsilon_1\cdot\psi(z_1)) 
	(\epsilon_2\cdot\psi(z_2)) 
	\prod_{i=3}^N :(k_i\cdot \psi(z_i))(\epsilon_i\cdot \psi): \rangle\,.
\end{equation}
The integral
\begin{equation}
	\int[d \psi]\epsilon_1\cdot\psi(z_1)) 
	(\epsilon_2\cdot\psi(z_2)) 
	\prod_{i=3}^N :(k_i\cdot \psi(z_i))(\epsilon_i\cdot \psi):
\exp\left({-1/2\int \psi \bar\partial
\psi}\right)\,,
\end{equation}
can therefore be written in terms of the following $(2N-2)\times (2N-2)$ matrix:
\begin{equation}
	M'= \begin{pmatrix}
		\,A & -C'^{\rm T}\\
		\,C' & B
		\end{pmatrix}\ ,
\label{e:pfaff-pure-fermions}
\end{equation}
composed of the block matrices $A$, $B$  given in~\eqref{e:pfaff-gauge} and $C'$ for which we have
\begin{equation}
	C'_{i,i} = 0\,,\quad C'_{ij} = \frac{\epsilon_i\cdot k_j}{z_i-z_j}\,,\qquad
i=1,2,...,N\,,\quad j=3,4,...,N\,,\quad j\neq i\,.
\label{e:block-matrices}
\end{equation}
These matrices are of sizes $(N-2)\times(N-2)$, $N\times N$ and $N\times(n-2)$,
respectively, because the vertex operators corresponding
to particles $1$ and $2$ do not have corresponding $k_i\cdot\psi$. 

This is not yet the Pfaffian of eq.~\eqref{e:pfp-def} 
because the matrix $C'$
has $0$'s on the diagonal since the self contraction $\langle:(k_i\cdot
\psi(z_i) )(\epsilon_i \cdot \psi(z_i)):\rangle$ vanishes. 
This self contraction must be replaced by the bosonic contraction of a $\partial X$ field 
with the plane-wave factor just as in ref.~\cite[eq.~(6.2.25)]{Polchinski:1998rq} (see also~\cite{Mason:2013sva,Mafra:2011nv}),
\begin{equation}
	:(\epsilon_i\cdot \partial X(z_i)) e^{i \sum_{l} k_l X(z_l)}: 
	\,\sim \left(-2\alpha'\sum_{l}\frac{\epsilon_i\cdot
k_l}{z_i-z_l}\right)
	:e^{i \sum_{l\neq i} k_l X(z_l)}: + O(z_i-z_l)\,,
\end{equation}
providing the correct factor to add to the diagonal of the matrix $C'$
\begin{equation}
	C_{i,i} = -\sum_{l}\frac{\epsilon_i\cdot k_l}{z_i-z_l}\,,\quad
C_{ij}=C'_{ij}\,,\quad j\neq i\,,
\end{equation}
and thus matching the Pfaffian of the matrix $\Psi_{12}^{12}$. After including the
superghost correlator $\langle e^{\varphi_1} e^{\varphi_2} \rangle=
z_{12}^{-1}$, we end up with ${\rm Pf}' \Psi$ defined in
\eqref{e:pfp-def}.

In the approach of ref.~\cite{Mason:2013sva} 
there are here no other contractions to perform because the $\partial X$ field 
by construction is taken to be a momentum $P$ field frozen by the scattering
equations. However, here the story is different as we are here dealing with 
{ actual string theory}. The $\partial X$
fields do have nonvanishing OPEs with other $\partial X$ fields. This is
also the mechanism that prevents unwanted tachyon poles from appearing
in the string theory amplitudes.

In order to derive these remaining terms, one can simply
recursively apply Wick's theorem. 
In the first step, one finds OPEs only between $\partial X$'s
and the plane-wave factor; this gives the Pfaffian in eq.~\eqref{e:pfp-def}. 
In the second step, one performs all possible contractions between only two
$\partial X$'s, the rest as before; this yields a
sum of Pfaffians where two more sets of lines and rows have been crossed
out, with a corresponding $\langle \partial X(z) \partial
X(w)\rangle\sim(z-w)^{-2}$ propagator in front of it (this induces a weighing
$1/\alpha'$ compared to the term of the first step).
By iterating the process, one finally deduces that the chiral kinematic
correlator is expressed as a sum of Pfaffians and the full answer is
\begin{equation}
\begin{split}
	\cA^{}_N
	= \left(g_o\over\sqrt{\alpha'}\right)^{N-2}\, {\alpha'}^{N-3}\,\int \prod_{i=2}^{N-2}d z_i
	\prod_{1\leq i<j\leq N} & |z_{ij}|^{2\alpha' k_i\cdot k_j}\times(z_1-z_{N-1})(z_{N-1}-z_N)(z_N-z_1) \times\\
	\biggr(
	{\rm Pf'}(\Psi)+
	\sum_{k=1}^{\lfloor \frac N 2 \rfloor} \frac{1}{(2\alpha')^k} &
	\!\!\!
	\sum_{\text{distinct~pairs}\atop (i_3,i_4),...,(i_{2k-1},i_{2k})}
	\prod_{p=3}^{2k-1}
	\frac
		{(\epsilon_{i_p}\cdot \epsilon_{i_{p+1}})}
		{(z_{i_p i_{p+1}})^2}
	{\rm Pf\,}'(\Psi^{i_3i_4...i_{2k}}_{i_3i_4...i_{2k}})\biggr)\,,
\label{e:poly-pfaff}
\end{split}
\end{equation}
where $z_{ij}=z_i-z_j$ and a global normalization factor has been set to $1$ and where 
${\rm Pf}'(\Psi^{i_3i_4...i_{2k}}_{i_3i_4...i_{2k}})$ stands for
$\frac{1}{z_{12}}{\rm Pf}(\Psi^{12i_3i_4...i_{2k}}_{12i_3i_4...i_{2k}})$.

\section{From String Theory to Scattering Equations}\label{sec:IBP} 
In the previous section we have identified which piece of string theory gives
rise to the Pfaffian of eq.~\eqref{e:pfp-def}, and which yields
additional terms.  We will now show that the additional terms, through
partial integrations, can be put in a form that makes them proportional
to the scattering equations, causing them to vanish with the
alternative integration measure that imposes scattering equations as a
delta function constraint.  In this form the full expression can be
integrated over these two different measures, both yielding the
correct field theory result when taking the $\alpha' \to 0$ limit.
Some simple examples will illustrate this.\\
 
Let us for simplicity focus first on the four-gluon amplitude. As
explained in the previous section, it takes the form
\begin{equation}\label{e:A4point}
\mathcal	A_4(1,2,3,4)	= \left(g_o\over\sqrt{\alpha'}\right)^2\,{\alpha'}\,\int_0^1 \left({\rm Pf}'( \Psi)+ 
	\frac{({\epsilon_1}\cnull {\epsilon_2})({\epsilon_3}\cnull {\epsilon_4})}
		{2\alpha' z_2^2}\right)\,
	z_2^{2\alpha'k_1\cdot k_2}(1-z_2)^{2\alpha'k_2\cdot k_3}\, dz_2\,,
\end{equation} 
where as usual $z_1=0$,  $z_3=1$, and $z_4=\infty$.
The additional piece proportional to $1/\alpha'$ is crucial in the string theory
context, as it removes a tachyon pole and allows the limit $\alpha' \to 0$ to be 
taken, yielding the field theory answer. 
 
One notices that the term 
\begin{equation} 
\delta \mathcal A_4= \int_0^1 dz_2\,\frac{1}{z_2^2}\exp\big(2 \alpha' k_1\cdot k_2  
\log(z_2)+2\alpha' k_2\cdot k_3  \log(1-z_2)\big)\,, 
\label{alphap}\end{equation} 
can be integrated by part to give
\begin{eqnarray} 
\delta \mathcal A_4&=&-\int_0^1 dz_2\,\partial_{z_2}\left(\frac{1}{z_2}\right)\exp\big(2
\alpha' k_1\cdot k_2  \log(z_2)+2\alpha' k_2\cdot k_3
\log(1-z_2)\big)\cr
&=&
\int_0^1 dz_2\,\frac{1}{z_2}\,\partial_{z_2}\Big(\exp\big(2 \alpha' k_1\cdot k_2 \log(z_2)+2 \alpha' k_2\cdot k_3 \alpha' \log(1-z_2)\big)\Big) \,. \ \
\label{IBP4} 
\end{eqnarray} 
By analytic continuation we can choose a kinematic region where  the boundary terms vanish.
Eq.~\eqref{IBP4} can be rewritten as
\begin{equation} 
\delta \mathcal A_4=\alpha'\,\int_0^1 dz_2\,\frac{1}{z_2}\Big(\frac{k_1\cdot k_2}{z_2} + \frac{k_2\cdot k_3}{1-z_2}\Big)\Big(\exp\big(2 \alpha' k_1\cdot k_2 \log(z_2)+2 \alpha' k_2\cdot k_3 \alpha' \log(1-z_2)\big)\Big) \,, \ \ 
\end{equation} 
where we recognize the four-point scattering equation
\begin{equation}
S_2 = \frac{k_1\cdot k_2}{z_2} + \frac{k_2\cdot k_3}{1-z_2} ~.
\end{equation} 
From this we can write new dual model prescription for gauge field
amplitudes by evaluating the string integrand on the solution of the
scattering equation by inserting the delta function factor given in~\eqref{deltafunction}. Since the  $1/\alpha'$ term is proportional to the
scattering equation in~\eqref{e:A4point} we have (using the relation
between the open-string coupling constant the 
 Yang-Mills coupling constant in ten dimensions $g_{o}=g_{\rm
  YM} \sqrt{\alpha'}$)
\begin{equation}
  \mathfrak A_4(1,2,3,4)= g_{\rm YM}^2\,\int_0^1\, {\rm Pf}'( \Psi)\,
	z_2^{2\alpha'k_1\cdot k_2-1}(1-z_2)^{2\alpha'k_2\cdot k_3-1}\,  \delta(S_2)  \, dz_2\,.  
\end{equation}

Another ordering of the external legs will yield another scattering
equation.  The various ordered amplitudes are of course related by the
action of the momentum kernel~\cite{BjerrumBohr:2010hn}. 
 
We see that in string theory we can trade the explicit $1/\alpha'$ term
by an integration over a term proportional to the scattering equation. In
string theory this term of course gives a contribution. 

The same phenomenon occurs for amplitudes with higher $N$. It gets increasingly
tedious to carry out the sequence of partial integrations, but the origin
of the mechanism seems to be closely related to a similar situation in
string-based rules, proven in Appendix B of ref.~\cite{Bern:1990ux} (see also ref.~\cite{Mafra:2011nv}).
In this procedure, the last
step is always a single integration by part on a variable that has been
isolated, which, when the partial derivative hits the Koba-Nielsen factor,
brings down a scattering equation in the integrand, just as in this four-point
example, leading the following form for the new dual model amplitude
prescription

\begin{multline}
	\mathfrak A_N(1,2,3,\dots,N)
	= g_{\rm YM}^{N-2}\,\int \prod_{i=2}^{N-2}d z_i
	\prod_{1\leq i<j\leq N} (z_i-z_j)^{2\alpha' k_i\cdot k_j}\times\cr
        {(z_1-z_{N-1})^2(z_{N-1}-z_N)^2(z_N-z_1)^2\over  \prod_{i=1}^{N}
          (z_i - z_{i+1})}	\times {\rm Pf'}(\Psi) \prod_{i\neq1,N-1,N}\delta(S_i)\,.
\label{e:gaugenewdual}
\end{multline}

After having done these partial integrations, the new integrand now
has the property that it corresponds to the CHY integrand at first
order in $1/\alpha'$. As we already emphasized, this is natural from
the point of view of the ambitwistor string models
\cite{Mason:2013sva,Adamo:2013tsa}.

Once again, the reason for this is because we have shown that the
higher order term in $1/\alpha'$, after IBP reduction, is exactly
killed by the scattering equation constraint. Although we calculate the Pfaffian
according to standard conformal field theory rules, the integrations
by part of the $1/\alpha'$-terms are only a valid operation in the
string theory integrand.  This is why one can set 
$\alpha'$ to zero in the integrand to recover the CHY prescription,
without meeting any singularities. This is very different from the
usual infinite tension limit of string theory where one needs to
scale the variables of integrations to reach the pinching limits of
the string integrand (see~\cite{Tourkine:2013rda} for a recent discussion). 

In the Gross-Mende $\alpha'\to\infty$ limit, the $1/\alpha'$
correction for the string amplitudes in~\eqref{e:poly-pfaff} vanish.
Consequently, the string theory amplitude and the new dual model
prescription in~\eqref{e:gaugenewdual} have 
the $\alpha' \to \infty$ Gross-Mende saddle point, but with different prefactors compared
to the usual high-energy limit of the string theory amplitudes. 

 \section{Amplitudes with Fermions and Mixed Amplitudes}\label{sec:fermions}

\subsection{The Four-Fermion Amplitude}
In this section we show the generality of the delta function measure (\ref{deltafunction}) by
calculating a few tree level 
amplitudes directly from string theory integrands. As a first example,
we check how fermion amplitudes can come out from our prescription. 
In the case of the fermion four-point amplitude one has~\cite{Cohn:1986bn,fms}
\begin{equation}
{\mathcal A}_4 = \left(g_o\over\sqrt{\alpha'}\right)^2\alpha'\,\int_0^1 dz_2 z_2^{-2\alpha't -
  1}(1-z_2)^{-2\alpha's -1}\left[(1-z_2)(\bar v_1\gamma^{\mu}u_2)
(\bar v_3\gamma_{\mu}u_4)- z_2(\bar v_1\gamma^{\mu} u_4)(\bar v_3\gamma_{\mu}u_2)\right]\,,
\end{equation}
where $\bar v_i$ and $u_i$ are the incoming and outgoing fermion wave functions.
As is well known, this string theory integral can be done in terms of two beta functions. 
In the field theory limit $\alpha' \to 0$ it of course yields the correct answer
corresponding to the two channels $s$ and $t$.

But this integral also defines the correct field theory limit if we instead
integrate over the delta function measure given by the scattering equations
as provided by the additional measure factor (\ref{deltafunction}),
\begin{equation}\begin{split}
{\mathfrak A}_4 &= g_{\rm YM}^2 \int_0^1 dz_2\, \delta\left(S_2 \right)z_2^{-2\alpha't - 2}(1-z_2)^{-2\alpha's -2}\times \\ &\hskip5cm
\left((1-z_2)(\bar v_1\gamma^{\mu} u_2)(\bar
   v_3\gamma_{\mu} u_4) - z_2(\bar v_1\gamma^{\mu} u_4)(\bar v_3\gamma_{\mu}u_2)\right)\,,
\end{split}\end{equation}
where $S_{2}$ is the scattering equation in $k_2$.
Explicitly, we get in the limit $\alpha' \to 0$, 
\begin{equation}
 A_4 = g_{\rm YM}^2\left[\frac{1}{s}(\bar v_1\gamma^{\mu} u_2)(\bar
   v_3\gamma_{\mu} u_4)
- \frac{1}{t}(\bar v_1\gamma^{\mu} u_4)(\bar v_3\gamma_{\mu}u_2)\right]\,,
\end{equation}
which is the correct field theory answer. 

\subsection{The Two-Fermion Two-Gluon Amplitude}
As another example of how this procedure works, one can similarly 
work out the expression for the two-fermion two-gluon amplitude. For the corresponding
string theory integrand see, $e.g.$, refs.~\cite{Cohn:1986bn,fms}. This amplitude has also been 
considered in the ambitwistor framework of ref.~\cite{Adamo:2013tsa},
but here we explain how to derive the result starting from  ordinary string theory.

We have explicitly verified in this case that the delta function 
measure (\ref{deltafunction})
yields exactly the tree level amplitude in the limit $\alpha' \to 0$. In this
case it follows in essentially one line, as there are no cancellations between tachyonic terms in the 
amplitudes. 
It indeed seems that we can directly take superstring integrands for amplitudes including
fermions and
integrate over a measure that localizes exactly on the scattering equations.  

\subsection{The Five-Point Mixed Scalar-Gluon Amplitude}
To give further credence to the procedure, 
let us finally consider a five-point case involving mixed external states of four scalars and a gluon.
Because of the combination of scalars and a gluon, the string theory integrand of this amplitude
contains two tachyonic terms canceling each other in the integral, and we again first
make this cancellation manifest by means of a single partial integration. 
We borrow the expression for the string theory integrand of the amplitude from
ref.~\cite{Stieberger:2007jv} (the explicit prefactor $K_a$ in front of the integral can be 
found in that paper, but we do not need it for the arguments here),
\begin{equation}
\begin{split}
{\cal A}_5 (\phi_1,\phi_2,\phi_3,\phi_4, g_5)&= K_a
\int\Big(\prod_{k=4}^5 d z_k\Big)\Big(\prod_{i<j}z_{ij}^{\alpha' s_{ij}}\Big)
\Big(\frac{1}{z_{35}}\Big(\frac{(\zeta_5\cdot k_4)}{z_{45}}
\frac{\alpha' s_{12} z_{34}} {z_{24} z_{13} z_{14} z_{23}}\Big)\\ &\hskip-2cm
+\frac{(\zeta_5\cdot k_1)}{z_{15} z_{24}}\Big(\frac{(1-\alpha' s_{24})}{z_{24}z_{13}}
+\frac{\alpha' s_{24}}{z_{14}z_{23}}\Big)+
\frac{(\zeta_5\cdot k_2)}{z_{14}z_{25}}\Big(\frac{(1-\alpha' s_{14})}{z_{14}z_{23}}+
\frac{\alpha' s_{14}}{z_{13}z_{24}}\Big)
\Big)\,,
\end{split}
\end{equation}
where $s_{ij}=2k_i\cdot k_j$, $\zeta_i$ and $k_i$ are the polarizations and momenta. 
Using the integration-by-parts relation in $z_4$ for the terms with $\zeta_5$ dotted 
with $k_1$ and $k_2$ 
we can rewrite these explicit $1/\alpha'$ terms exactly as in the pure gluon case.
This replaces that term by the scattering equation in leg 4, {\it e.g.} 
$(1- \alpha'  s_{14})\to({\rm IBP}(S)_4 z_{14}-\alpha' s_{14})$ . 
Using the
prescription (\ref{deltafunction}), we get
\begin{equation}
\begin{split}
{\mathfrak A}_5 (\phi_1,\phi_2,\phi_3,\phi_4, g_5)&= K_a
\int\Big(\prod_{k=4}^5 d z_k\Big)\Big(\prod_{i<j}z_{ij}^{\alpha' s_{ij}}\Big)
\delta\left(S_4\right)\delta\left(S_5\right)\frac{z_{12}^2z_{23}^2z_{31}^2}{\prod_{1\leq
    i \leq 5} (z_i-z_{i+1})}\\ &\hskip-2.5cm
\Big(\frac{1}{z_{35}}\Big(\frac{(\zeta_5\cdot k_4)}{z_{45}}
\frac{\alpha' s_{12} z_{34}} {z_{24} z_{13} z_{14} z_{23}}\Big)
+\frac{(\zeta_5\cdot k_1)}{z_{15} z_{24}}\Big(\frac{\alpha' z_{24}S_4-\alpha' s_{24}}{z_{24}z_{13}}
+\frac{\alpha' s_{24}}{z_{14}z_{23}}\Big)\\ &\hskip3cm+
\frac{(\zeta_5\cdot k_2)}{z_{14}z_{25}}\Big(\frac{\alpha' z_{14}S_4-\alpha' s_{14}}{z_{14}z_{23}}+
\frac{\alpha' s_{14}}{z_{13}z_{24}}\Big)
\Big)\,,
\end{split}
\end{equation}
where the delta function measure now has been adapted to the situation where legs $(1, 2, 3)$ are fixed as 
$(-\infty,0,1)$ following the convention used in in~\cite{Stieberger:2007jv}.
We see that the delta function effectively removes the $1/\alpha'$ term after having 
canceled the tachyon pole explicitly by use of the partial integration that introduces
the scattering equation in leg 4, $ S_4=0$. After some algebra we 
arrive in the limit of $(\alpha'\to 0)$ at
\begin{equation}
{ A}_5 (\phi_1,\phi_2,\phi_3,\phi_4, g_5)=K_{ft}\Big(
(\zeta_5\cdot k_1)\Big(\frac{1}{s_{23}}-\frac{s_{34}}{s_{23}s_{15}}\Big) + (\zeta_5\cdot k_2)\Big(\frac{1}{s_{23}} \big)+ (\zeta_5\cdot k_4)\Big(\frac{s_{12}}{s_{23}s_{45}}\Big)\Big)\,,
\end{equation}
which is the correct result. Here $K_{ft}$ denotes the prefactor of
the amplitude in the limit $(\alpha'\to 0)$.  There thus seem to be no
additional problems associated with mixed amplitudes. We therefore
expect that any generic amplitude involving gluons, scalars and
fermions in any combination can be computed in the same manner,
imposing the same delta function measure after having manifestly
canceled all tachyon poles (if present) through integrations by
parts.
 
\section{Conclusion}\label{sec:conclusion}
\noindent
We have provided an natural interpolation between the CHY prescription
for tree level amplitudes in field theory and the Gross-Mende limit of
string theory. We have introduced a new kind of dual
model defined as the string theory localized on the surface of the
solutions to the scattering equations.
We have shown how this can be used to derive new amplitudes, those with external
fermions, on the basis of merging string theory with the scattering
equations. Numerous other examples can be derived similarly: mixed
amplitudes with gluons and fermions, scalars and fermions, and so
on. We have provided some examples, and argued that the general
prescription is to rewrite the string integrand by manifestly canceling tachyon poles 
and then evaluate the string integrand on the solutions to the
scattering equations.  It would be very interesting to relate the
prescription given in this paper to an $\alpha'$-extension of
Berkovits' prescription given in~\cite{Berkovits:2013xba}.

From this prescription no further calculations are necessary since one
can use the form of the string integrand with the $1/\alpha'$
expansion of Pfaffian, perform the partial integration to remove the
second order poles, and evaluate it on the scattering equations.

Another very important question concerns closed string. 
The whole CHY construction, and the subsequent ambitwistor/pure
spinor models are intrinsically closed-string like
models. The way in which scalar, gauge, or gravity interactions are
implemented at the integrand level is indeed highly reminiscent of the
string theory realization of these interactions, by the left-right
moving mixing \cite{Ochirov:2013xba}. It is an interesting question
how the prescription used here transcribe into closed-string language.
The reason this is nontrivial is the absence of chirality in the
closed string, where one sector is holomorphic while the other is
antiholomorphic. This is very different from the CHY prescription,
where both sectors of the theory possess the same chirality as in ambitwistor 
models.  

An obvious question is what happens at the genuine quantum level,
$i.e.$ at loop order.  Tree level amplitudes correspond to vertex
operators on the sphere.  Using again string theory as the guide, one
would be led to consider the corresponding scattering equations
associated with the $N$ external momenta but integrated over
correlation functions on higher genus surfaces. Integrations will
remain even after imposing the scattering equations. It would be
interesting to see if they reproduce the result of field theory loop
computations in the $\alpha' \to 0$ limit.

\section*{Acknowledgements}\label{sec:Acknowledgements} 
We thank Paolo Di Vecchia for useful 
comments. We also thank David Fairlie 
for sending us a copy of his unpublished paper together with 
D.~E.~Roberts, ref.~\cite{Fairlie:1972zz}. 
We acknowledge support from the Agence Nationale pour la Recherche (ANR) grant  reference 
QFT ANR 12 BS05 003 01, and Projet International de Coopération Scientifique (PICS)  grant 6076.

\appendix
\section{Appendix: Further details on integration by parts.}

In Section 5 we did not want to clutter the text with more explicit details of higher-point
issues with respect to the needed integration by parts. In this Appendix we provide
a few details of what happens at five points.

To illustrate in this slightly more complicated case
how to do the integration by parts, we consider the term
originating 
from the $\partial X(z_i) \partial X(z_j)$ contractions in the ghost picture changing
formalism. As in the four-point case we choose to remove lines 1 and 2
in the matrix of the Pfaffian.
 
Explicitly, we have the following two types of terms that are of type
$\frac{1}{\alpha'}$ 
\begin{equation}\begin{split} 
\sim \ldots -\frac{\epsilon_1\cdot k_3 \epsilon_2\cdot\epsilon_3
\epsilon_4\cdot\epsilon_5}{ \alpha'  z_{12} z_{13} z_{23} z_{45}^2} 
 +\frac{\epsilon_1\cdot\epsilon_3 \epsilon_2\cdot k_3
\epsilon_4\cdot\epsilon_5}{ \alpha'  z_{12}z_{13} z_{23} z_{45}^2} 
 -\frac{\epsilon_1\cdot k_4 \epsilon_2\cdot\epsilon_4
\epsilon_3\cdot\epsilon_5}{ \alpha'  z_{12} z_{14} z_{24} z_{35}^2}\\ 
 +\frac{\epsilon_1\cdot\epsilon_4 \epsilon_2\cdot k_4\epsilon_3\cdot\epsilon_5}{
\alpha'  z_{12} z_{14} z_{24} z_{35}^2} 
 -\frac{\epsilon_1\cdot k_5 \epsilon_2\cdot\epsilon_5
\epsilon_3\cdot\epsilon_4}{ \alpha'  z_{12} z_{15} z_{25}z_{34}^2} 
 +\frac{\epsilon_1\cdot\epsilon_5 \epsilon_2\cdot k_5
\epsilon_3\cdot\epsilon_4}{ \alpha'  z_{12} z_{15} z_{25} z_{34}^2}\,, 
  \label{firsttype5} \end{split}\end{equation} 
 and  
 \begin{equation}\begin{split} \sim& \ldots
 -\frac{\epsilon_1\cdot \epsilon_2 \epsilon_3\cdot \epsilon_4\left( 
   \frac{\epsilon_5\cdot k_1}{z_{15}} 
 +\frac{\epsilon_5\cdot k_2}{z_{25}} 
 +\frac{\epsilon_5\cdot k_3}{z_{35}} 
 +\frac{\epsilon_5\cdot k_4}{z_{45}}\right)}{ \alpha'  z_{12}^2 z_{34}^2}
\\&\hskip-0.7cm
 -\frac{\epsilon_1\cdot \epsilon_2 \epsilon_3\cdot \epsilon_5\left( 
 \frac{\epsilon_4\cdot k_1}{z_{14}} 
+\frac{\epsilon_4\cdot k_2}{z_{24}} 
+\frac{\epsilon_4\cdot k_3}{z_{34}} 
-\frac{\epsilon_4\cdot k_5}{z_{45}}\right)}{ \alpha'  z_{12}^2 z_{35}^2}
+\frac{\epsilon_1\cdot \epsilon_2 \epsilon_4\cdot \epsilon_5\left( 
-\frac{\epsilon_3\cdot k_1}{z_{13}} 
-\frac{\epsilon_3\cdot k_2}{z_{23}} 
+\frac{\epsilon_3\cdot k_4}{z_{34}} 
+\frac{\epsilon_3\cdot k_5}{z_{35}}\right)}{ \alpha  z_{12}^2 z_{45}^2}\,.
\label{secondtype5}\end{split}\end{equation} 
 We will now show that in all cases we can find integration-by-part
relations that are
equivalent to inserting the scattering equations.  
 \begin{itemize}
\item In the first equation~\eqref{firsttype5} we will in terms 1 - 2 use
the relation involving $z_4$, while for the 
terms 3 - 6 we will instead use the integration-by-part relation in $z_3$. 
 
\item In the second equation~\eqref{secondtype5} for the first term we will use
the relation in the variable 
$z_3$, except for the next-to-last term where we will use the one for $z_4$. For the
second and third terms here we will use those in $z_1$, except
for the first terms where we use those in $z_3$ and $z_4$.
\end{itemize}

By this prescription we have absorbed all the $\frac{1}{\alpha'}$ terms of
the five-point amplitude. Again
we observe that at the solution to the scattering equations the reduced Pfaffian
will be unchanged, since all we have done is to turn them into terms proportional to the
scattering equations.


\end{document}